# Abstracting spreadsheet data flow through hypergraph redrawing

David Birch[1], Nicolai Stawinoga*[1], Jack Binks[2], Bruno Nicoletti[2], Paul Kelly[1]
[1]Imperial College London
[2]Filigree Technologies, London
* nicolai.stawinoga@imperial.ac.uk

**ABSTRACT**

We believe the error prone nature of traditional spreadsheets is due to their low level of abstraction. End user programmers are forced to construct their data models from low level cells which we define as "a data container or manipulator linked by user-intent to model their world and positioned to reflect its structure". Spreadsheet cells are limited in what they may contain (scalar values) and the links between them are inherently hidden. This paper proposes a method of raising the level of abstraction of spreadsheets by "redrawing the boundary" of the cell.

To expose the hidden linkage structure we transform spreadsheets into fine-grained graphs with operators and values as nodes. "cells" are then represented as hypergraph edges by drawing a boundary "wall" around a set of operator/data nodes. To extend what cells may contain and to create a higher level model of the spreadsheet we propose that researchers should seek techniques to redraw these boundaries to create higher level "cells" which will more faithfully represent the end-user's real world/mental model. We illustrate this approach via common sub-expression identification and the application of sub-tree isomorphisms for the detection of vector (array) operations.

## 1 INTRODUCTION

Spreadsheets are known to be the most common form of end user programming. This likely makes the spreadsheets end users create the most common software representation of end users' mental model of their modelling and data manipulation problems. For example a user may model the growth of capital in a savings account using a simple compound interest spreadsheet model.

As discussed by (Hutchins, et al., 1985) users iteratively build improved mental models of how a system works in order to bridge both the "gulf" of evaluation and of understanding. Users first encounter a "gulf evaluation" in understanding the current state of the a system such as a spreadsheet which may be held in hidden formulas. Users then face a "gulf of execution" in understanding how to impact the system as they desire. However even if the user build a mental model that mirrors exactly how the spreadsheet functions we must be aware that many spreadsheets do not quite function as their creators intended (not everyone understands compound interest for example!). Indeed spreadsheets are well known to be error prone(Panko, 1998) and regularly underly large financial mishaps. Aside from human nature many reasons are posited for this in relation to the underlying spreadsheet model and its implementation in popular software packages such as Microsoft Excel or Google Sheets.

In this paper we consider one aspect of these challenges namely the low level of abstraction within a spreadsheet. In creating spreadsheet models users are forced to map their data and it's manipulation into a rectangular cell grid. This means that lists, tables, hierarchies and other multidimensional data structures are split apart into cells only related by their locality and possibly by visual formatting cues. Such entities may no longer be manipulated conceptually as a high level entity, as a user would naturally think of doing ("Sum sales of Widgets for 2016"). Instead such manipulations are expressed as multiple operations on collections of atomised cells. This is known to be an error prone process.

Coupled with this low level of abstraction is the absence of any higher level of modelling abstraction which would be present in most other programming environments (structures, enumerations, objects, class hierarchies, UML diagrams …). The absence of a formal higher level description of the user's model precludes model verification – that is ensuring that the model implementation is correct "are we building the right model". More importantly it makes the validation of models difficult – that is checking if "we are building the right model"(Easterbrook, 2010). This is particularly challenging when users are presented with a spreadsheet they did not create as they must build a mental model of the purpose of the spreadsheet from its atomised low-level primitives.

Unfortunately few tools exist (see Section 10) to express a higher level of abstraction in spreadsheet models. Further we find that few tools or methods exist to discover a higher level model description from the spreadsheet.

In this paper we seek to introduce and illustrate a method of creating higher level abstractions for spreadsheet models.

## 2 PROBLEM

We believe that one of the key reasons for the challenges encountered with spreadsheets is their low level of abstraction. This abstraction level rests firmly at the level of cells which for the purpose of this paper we define as:

*"a data container or data manipulator*

*linked by user-intent to model their world*

*and positioned to reflect its structure"*

Cells contain single data values such as numbers, text or dates or the formula to calculate such a value. These calculations may reference other cell values or groups of cell values. These links form the a calculation structure which builds a model of the user's world. For example as in Figure 1 summing the sales of all sales agents and calculating their sales bonuses. In addition we see that the positioning of cells within the grid structure is generally done to reflect higher level structures in the real world – e.g. a list sales.

|   | A | B | C | D |   |   | A | B | C | D |
|---|---|---|---|---|---|---|---|---|---|---|
| 1 | Agent | Sales (sqft) | Sales(m2) | Bonus |   | 1 | Agent | Sales (sqft) | Sales(m2) | Bonus |
| 2 | Fred | 5221 | 485.2952 | 100 |   | 2 | Fred | 5221 | =B2/3.28/3.28 | =IF(C2>AVERAGE(C$2:C$4),100,0) |
| 3 | Dave | 3872 | 359.9048 | 0 |   | 3 | Dave | 3872 | =B3/3.28/3.28 | =IF(C3>AVERAGE(C$2:C$4),100,0) |
| 4 | Bob | 3651 | 339.3627 | 0 |   | 4 | Bob | 3651 | =B4/3.28/3.28 | =IF(C4>AVERAGE(C$2:C$4),100,0) |
| 5 |   |   |   |   |   | 5 |   |   |   |   |
| 6 |   | Total Sales | 1184.563 |   |   | 6 |   | Total Sales | =SUM(C2:C4) |   |
| 7 |   | Total Bonus | 100 |   |   | 7 |   | Total Bonus | =SUM(D2:D4) |   |

*Figure 1 Simple spreadsheet to calculate total sales and bonuses for agents. Note the repetitive formulas.*

This definition leads us to highlight to problems with current implementations of cells which have bearing directly upon the low level of abstraction of the spreadsheet.

Firstly we see that spreadsheet cells contain a single value rather than a conceptual entity such as a list of products, this creates fragmentation of concepts into a collection of cells which are grouped only by their collocation in a grid, their adjacent labelling or formatting or by being commonly referenced as a group.

Consequently we see that all data manipulations (excluding array or table formulae) must again produce a single value such that conceptual operations must also be atomised rather than expressed concisely ("let the sales target for next year be 10% higher than this year's sales")

Secondly we see that the links between cells are normally hidden and difficult to comprehend. End users must interrogate individual cells to identify what references they make and how they link with other cells to contribute their calculation to the users model. While some tools exist to expose this structure(Hermans, et al., 2011), it remains inherently hidden. This makes understanding the model structure difficult.

Thirdly we see that the higher level structure of a model is only expressed informally through the positional structure of the spreadsheet. Generally we find that adjacent cells are conceptually related with some forming table dimensions (e.g. time expressed as Q1,Q2,Q3..) some forming the names of cells (e.g. "Profit:"). By this we find that cells may form lists, tables and dimensions to represent higher level concepts which model the structure of the user's world.

Finally we note that these encumbrances mean that the manipulation of the abstract concepts within a spreadsheet is challenging as not one cell but many together represent a higher level concept. Unfortunately few tools are available to manipulate groups of cells effectively without large amounts of manual work, for a good example see(Sarkar, et al., 2018).

As such we believe that while the concept of a cell remains effective there is a pressing need to augment the current implementation of a cell which restricts the level of modelling abstraction to be too low and difficult to manipulate.

## 3    THE IDEA

In this paper we propose that one way to achieve a higher level of modelling abstraction is to expand what constitutes a cell. We believe the best method to do this is to start from existing spreadsheet cells and manipulate them in such a way as to both raise the abstraction level and provide users with a more faithful representation of their model.

Conceptually we undertake this by "redrawing the boundaries" of what constitutes a "cell". We propose this by the following general steps:

1) First we seek to make the link structure explicit by transforming the spreadsheet into a graph structure. The reference structure now exposed should reflect the model of the users intent. In contrast to previous approaches (discussed in related work) we generate a fine-grained graph with functions, operators and values as nodes.
2) We then consider this graph as a hyper-graph, that is a graph where edges may connect not two but a set of any number of nodes. We observe that under this conceptualisation Cells may be represented precisely as hypergraph edges - that is a linked set of function, operator, value nodes. Visually one may express a hyper edge for a cell by drawing a boundary line or "cell wall" around a group of function/operator/value nodes. An example may be seen in Figure 7.
3) In order to raise the level of abstraction in spreadsheets we propose that researchers should seek techniques to redraw these boundaries of what constitutes a cell. Such "cells" would then provide end users with a better building block to more faithfully represent their world.

To illustrate this concept this paper seeks to use this approach to detect vector (array) operations using isomorphic subtree detection. Conceptually this algorithm seeks to identify "identical" cells and group them into higher level vector operations.

We demonstrate the generality of this approach via the detection of common sub tree expressions across the whole spreadsheet graph. These may represent unit conversions or other repeatedly used calculation.

We report the application of this method to the Enron spreadsheet corpus and are able to report a reduction in model complexity (as measured by cell count) of 27% via the identification of input vectors and vector operations.

We believe this graph-driven approach provides a mechanism for a wide range of spreadsheet analysis and refactoring tools to be constructed and many recently proposed tools could be re-implemented using this approach which we commend to the community via our open source toolkit.

We illustrate the effectiveness of raising the abstraction level by translating a series of spreadsheets to a high level data-flow language called ModL which has much bigger "cell walls". Such models have a significantly reduced number of "cells" which are of a higher level of abstraction.

# 4 CONTRIBUTIONS:

We believe that the contributions of our work are as follows:

1. In section 5 we demonstrate how two force directed layout algorithms reveal hidden spreadsheet structure when applied to spreadsheet derived cell level graphs.

2. In section 6 we introduce a fine-grained operator-level graph representation of spreadsheets and demonstrate its versatility by constructing such a graph for each of 99+% of the Enron spreadsheet corpus.

3. We propose a general method for abstracting spreadsheet models by "redrawing the cell walls" when cells are represented as hyper-graph edges in a fine-grained graph representation of the spreadsheet, this is shown in section 7.

4. In section 8 we demonstrate this by introducing vector operation detection via sub-tree isomorphism and find a compression rate of 27% on a sample of 2124 Enron spreadsheets.

5. We further show the potential opportunity for common sub-expressions to be factored out of the spreadsheet hypergraph by highlight how prevalent structurally isomorphic expressions are within the Enron corpus.

6. Finally in section 9, we introduce a graph based data flow modelling in an application called Wire with "larger cell walls" which can raise the abstraction level by enabling "cells" to contain and operate on multi-dimensional vectors. Manual translation results in a 95-98% reduction in model complexity as measured by formula counts.

# 5 EXTRACTING GRAPH STRUCTURE

*"linked by user-intent to model their world"*

Inspired by a long history of graph based representation of spreadsheets (Pertti & Sajaniemi, 1991)(Igarashi, et al., 1998)we seek to expose the hidden linking structure of the spreadsheet. These links are formed from cells with formulas which reference other cells, or ranges. This creates a directed acycling graph. We believe this structure is formed by end users as a reflection of their model of their world. That is, the way cells or groups of cells are linked together are a reflection of the entities and their relationship in the users world. For example profit may be calculated by subtracting sales from the cost of doing business.

### 5.1. Cell level Graph

Formally we form the cell graph by the following worklist algorithm starting with a user selected final result cell:

1. Add first cell to the worklist
2. For each cell in the worklist:
3. Parse the cell formula
4. Identify references within the formula
5. Remove this cell from the worklist
6. Expand all range references and named ranges to their constituent cells, add to the worklist

This algorithm may be run in parallel. We then form the graph with nodes for cells, ranges and named ranges; the latter with edges to their constituent cells. We then add edges to the graph between cells and the cells and ranges they have been determined to reference. Graphs are written to the standard GraphML file format(Brandes, et al., 2001) which can encode properties of nodes and edges. We encode the cells location (sheet, row and column) as well as their values and formula. This graph is then used as the basis for visual analysis.

### 5.2 Graph layout for Visual Analytics

To make sense of a graph it is necessary to map the nodes into two dimensional (XY) space. The objective of this mapping is to permit the structure of the graph formed by its edges to become apparent. A common family of algorithms for such graph layout are force direct algorithms. These algorithms treat the layout of a visualisation as a particle physics problem. A common example of such a problem would be to predict the motion of one or more planets as they orbits a sun which can be complex as each particle (sun, planet) exerts a gravitational force upon every other body. These algorithms treat the graph nodes as a particles with forces acting on each node. These forces are allowed to interact for a period of time as the graph nodes move through 2d space. These forces are defined in a variety of ways which reflect the edge structure of the graph and permit it to influence the layout of the final graph. The intention being to find "structure from chaos".

**Force Atlas 2**

In principle these algorithms are highly computationally expensive as the forces acting on one node may in general act upon all others, however a number of optimised algorithms have been developed which approximate the solution of these physics problems. One efficient such algorithm is the Force Atlas 2 algorithm (Jacomy, et al., 2014).

Under this algorithm the following forces are defined:

- ➢ A "spring" force is defined for each edge which seeks to draw its two nodes closer together.
- ➢ A "repulsion" force acts between every pair of nodes which seeks to spread the nodes apart.
- ➢ A "gravity" force is placed in the centre of the 2d space which seeks to draw all nodes toward it. This acts as a gravity well and stops nodes from flying off into infinite space.

Solving the particle physical problem for node locations influenced by these forces enables, with suitable configuration of their relative strengths, an informative trade off of clarity and complexity in the layout. Under these circumstances structures should become apparent.

As an example we consider a small 2700 cell spreadsheet model developed in the construction industry to answer a commercial question and explore the scenarios and circumstances around it. We create the cell level graph as described above and show it in Figure 2. Within this figure we see large islands of structure represent calculation models, generally, but not universally, there is a correspondence between these islands and different worksheets since they tend to be broadly independent from one another. These can be highlighted by a change of colour mapping. The orphaned cells are unreferenced and usually blank cells or textual labels which surround the extracted models. Further structures apparent in Figure 2:

A. A dragged out chain of cells of the form A2=A1*1.10 used to roll a percentage increase over time.

B. A shared and commonly used numeric assumption in the model.

C. A triangular structure of cells calculating net present value of future profit. Each year along the arrow is further into the future and must be projected back one year further, giving the growing complexity of the calculation.

D. A comparison between two business cases. Note the two structurally identical calculation chains being compared differ only by different numeric business assumptions.

E. Examples of the business assumptions being frequently referenced in the calculation model.

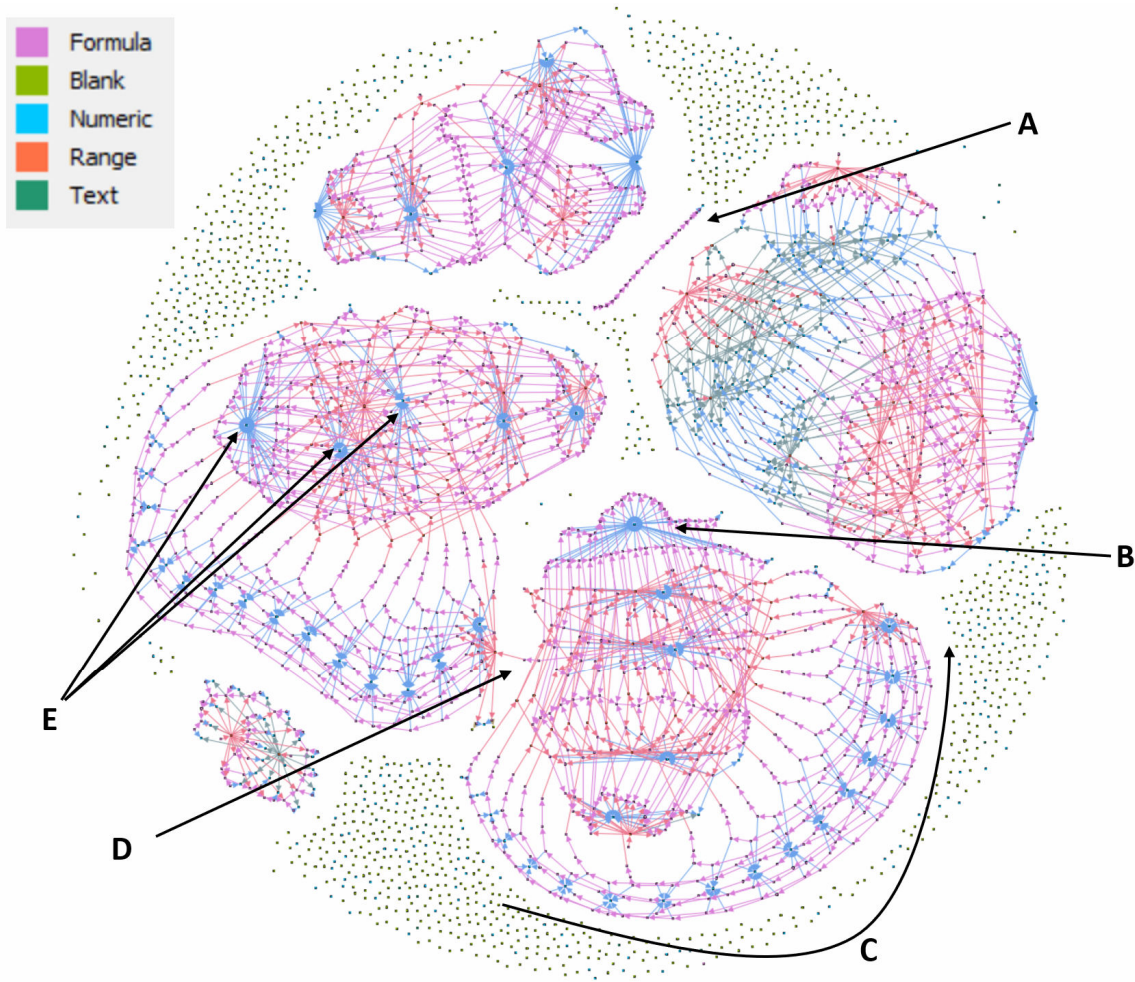

*Figure 2 Cell level graph of a typical model from the construction industry to answer a commercial question and exploring the effect of different build completion dates.*

**Multi-source of Gravity Force Atlas 2**

*"{and positioned to reflect its structure}"*

Developed by Suryansh Rastogi and David Birch the multi-source of gravity algorithm extends the Force Atlas 2 algorithm by enabling an infinite number of sources of gravity. Each node may be given its own source of gravity. This is done by specifying a gravity_x and gravity_y coordinate for each node. This enables a number of new possibilities for graph layout. Specifically for spreadsheets we may assign the gravity x and y to be the (integer) columns and row of each cell. For range and named range nodes we assign the top left value.

This permits the original spreadsheet grid placement structure to be reflected in the graph layout. Since the graph layout is solving a particle physics problem we may trade off the relative strengths of the various forces acting upon the nodes. Specifically we may trade the relative strengths of:

➢ The "spring" forces acting where edges are present to reflect the reference and dataflow structure of the spreadsheet

➢ The gravity forces which seek to tether nodes to their original grid locations.

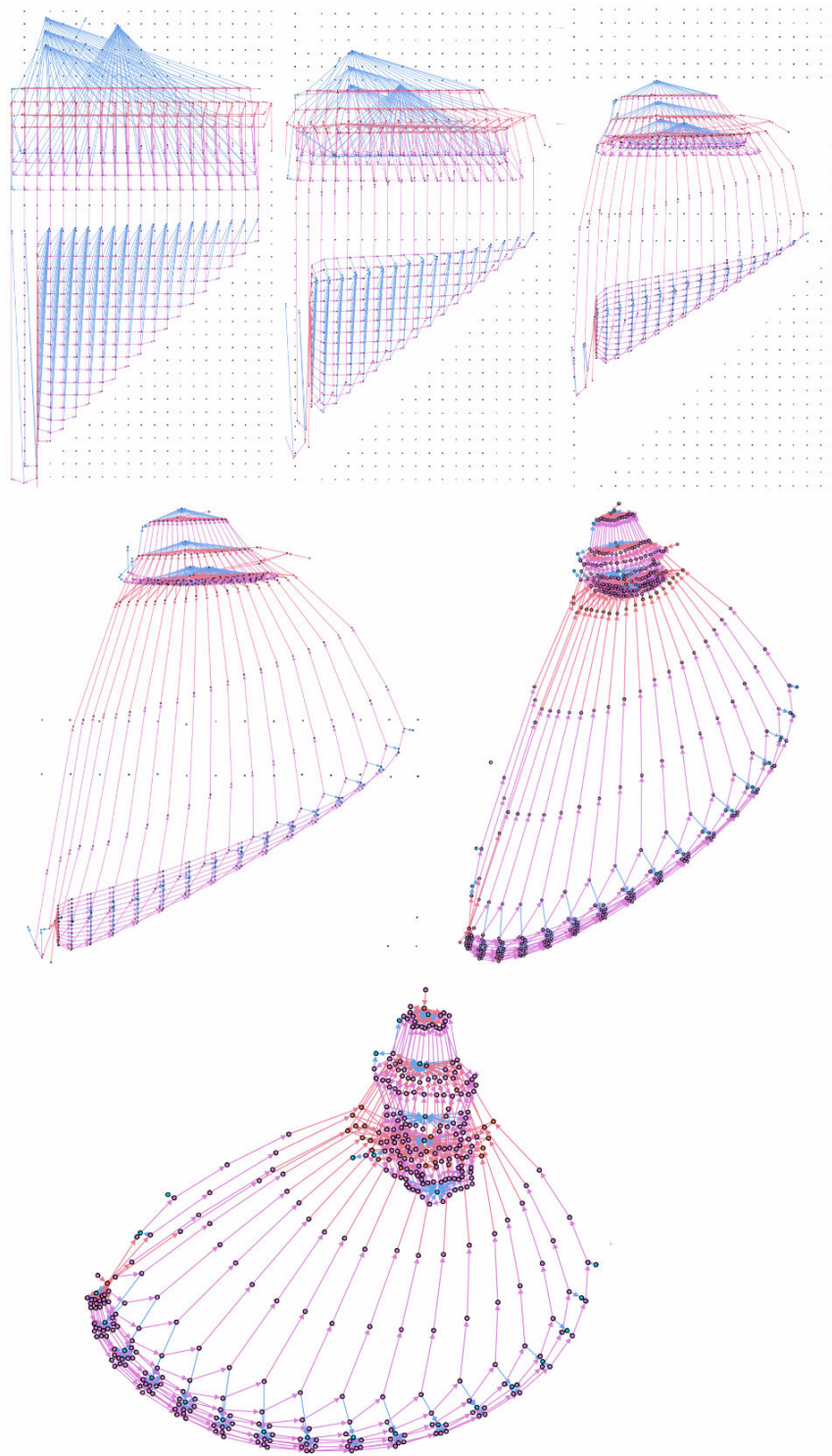

*Figure 3 Effect of varying the strength of gravity under the Multi-source of Gravity Force Atlas 2 layout algorithm. Note this is the same graph as in Figure 2. From top to bottom we see the gravitational force decreases relative to edge forces.*

Figure 3 shows the effect of changing the relative strength and interaction of these forces. The top left figure has gravity forces high enough to force all nodes to reside at their grid location. The triangular in the references which is given by the user is clearly apparent under high gravity conditions. This represents the triangular nature of the Net Present Value structure identified in Figure 2 C. As gravity decreases relative to the edge forces we see clusters of cells being to form each of which clearly depends upon the subsequent one – a structure not entirely clear in the triangular layout. In this way the increasing size and complexity of the group of cells calculating each year's net present value structure is clear from the growing clusters and the dependency year to year is also shown. Array (range) operations are also shown in the top half of each year with the dependency between them being clearest at intermediate levels of gravity.

As the strength of gravity is reduced through 5 orders of magnitude we see the grid structure weaken and enable the reference structure to determine, via its forces, the layout of the graph. This transition in force strengths enables a compromise to be sought which enables both the grid structure imposed by the user and the reference structure to be seen and learned from. This we believe will create the most informative graph layout in terms of identifying structure from both sources visually. In this example we believe the first graph of the second row has the best compromise of user dictated

Unfortunately it is currently unknown how to automatically balance these forces to find the most visually insightful graph and this must be done by human eye. Inspired by (Igarashi, et al., 1998) we would encourage the reader to experiment with the interactive variation of the relative balance of these forces particularly with sufficient computing power to enable interactive animation of the spreadsheet graph as it transitions between grid structure and reference structure. The graph layout is available as an open source plugin to the Gephi graph visualisation tool.

## 6   FINE LEVEL GRAPH

In contrast to the preceding graphs and the literature on graphs created from spreadsheet we now introduce a more fine-grained graph. Cell level graphs while they show structure still hide the calculation formulas, much as they are hidden within Microsoft Excel, requiring user interrogation to view.

To provide a graph view of the calculation structure inside cells we start with the parse tree graphs generated by the excellent XLParser (Aivaloglou, et al., 2015). This robust parser produces an Abstract Syntax Tree reflecting the formula structure with nodes representing values, functions and operators. Also included within these per-cell graphs are various annotation nodes which hint to an evaluator or printer how to parse or print the formula effectively.

Since this graph is complex and difficult to read we introduce 8 refactoring steps to reduce the verbosity of these syntax trees. These are shown in the table below and have the effect of turning the Parse tree into a clearer tree structure which operates from the root downward with the arguments for operators and functions forming their direct children.

| Refactoring | Purpose / Effect |
| --- | --- |
| `RemoveReferenceNodes-BeforeFunctionCalls` | Ensure that reference nodes appear only before cell references |
| `RemoveFormulaEqNode` | Remove the rootnode for formula trees so that the cell node may be used instead |
| `InlineFunctionNames` | Make the function name the root of the subtree |
| `RemoveConstantNodes` | Remove type indicator for constant nodes. |
| `RemoveFormulaNodes` | Make the function name the root of the subtree |
| `RemoveNumberNodes` | Remove type indicator for constant nodes. |
| `RemoveArgumentNodes` | Make the arguments of a node its direct children. |
| `TruncateReferences` | Truncate reference signatures for later processing. |

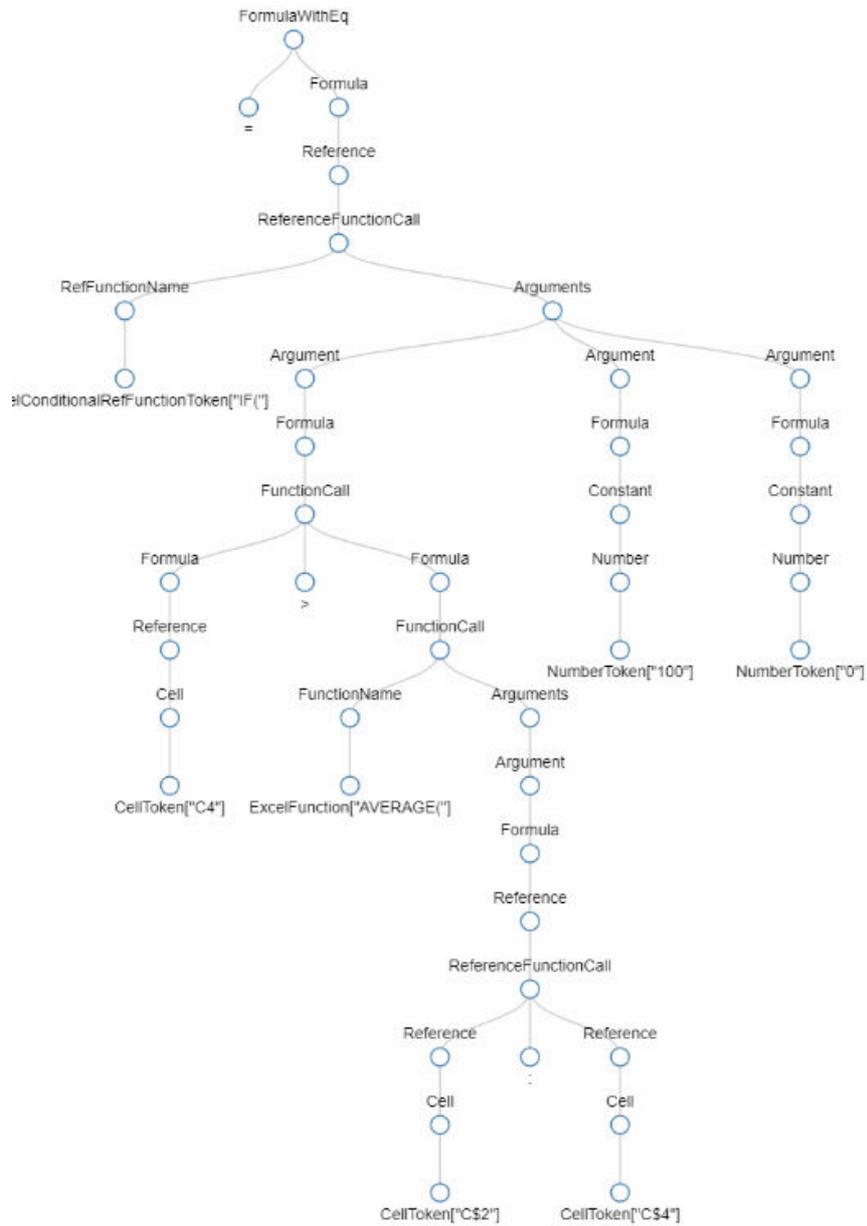

*Figure 4 Parse tree graph of the formula "=IF(C4>AVERAGE(C$2:C$4),100,0)" in cell D4 as generated by XLParser, prior to our refactoring. Compare with Figure 5 for the whole spreadsheet in Figure 5.*

*Figure 5 Fine-grained graph of the spreadsheet shown in Figure 1*

Thus far we have created a single cell's formula tree. While these are interesting we seek to form a whole spreadsheet fine-level parse tree led structure. This is done by expanding the Cell level graph discussed in section 5 by linking each cell node with its parse tree graph, such that each cell node links to the root of its parse tree. The edges originating from the cell node are now transferred to the relevant parse tree reference node which expresses the generated cell reference. Such references are identified by seeking Reference nodes in the formulas abstract syntax tree and converting their subtree into standard spreadsheet references to cells, ranges or named ranges. For the avoidance of doubt we continue to include explicit range and named range nodes which link to the cells they reference.

We further note that such graphs can be formed by extracting every cell within a spreadsheet rather than by recursively following references from one formula. The latter graphs form a sub-graph "slice" of a whole spreadsheet. For the remainder of the paper we now consider whole spreadsheet graphs.

To consider the generality of forming such complex graphs we sought to extract a fine-grained graph for each of the spreadsheets within the Enron spreadsheet corpus (Hermans & Murphy-Hill, 2015), of 15,399 spreadsheets considered 94.49% parsed, 4.72% failed due to external libraries providing a success rate of 99.2%. The remaining cases are caused by unusual reference structure. These graphs form the basis of our further analysis as we seek to raise the level of abstraction.

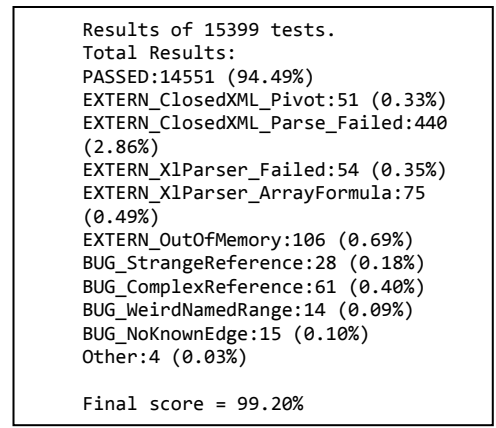

```
Results of 15399 tests.
Total Results:
PASSED:14551 (94.49%)
EXTERN_ClosedXML_Pivot:51 (0.33%)
EXTERN_ClosedXML_Parse_Failed:440
(2.86%)
EXTERN_XlParser_Failed:54 (0.35%)
EXTERN_XlParser_ArrayFormula:75
(0.49%)
EXTERN_OutOfMemory:106 (0.69%)
BUG_StrangeReference:28 (0.18%)
BUG_ComplexReference:61 (0.40%)
BUG_WeirdNamedRange:14 (0.09%)
BUG_NoKnownEdge:15 (0.10%)
Other:4 (0.03%)

Final score = 99.20%
```

*Figure 6 Results of generating fine-grained graphs for each of the Enron Corpus of Spreadsheets*

## 7 HYPERGRAPH VIEW

One challenge with the fine level graph is its complexity which hides the cell structure of the spreadsheet. To address this we can consider the graph as a hyper-graph, that is a graph where edges can connect not just two nodes but a set containing multiple nodes. Represented visually one can consider a hypergraph edge as drawing a boundary around a group of nodes which are contained within the hyperedge set.

An example of this hyper graph view of the graph may be found in Figure 7 which shows a hypergraph view of the fine grain graph shown in Figure 5.

Such a view provides both the high level cell-level graph discussed previously as well as the detailed structure of each cells data manipulation. Such graphs may be laid out recursively using either of the algorithms mentioned above at the cell level followed by a second application within each hypergraph edge. Alternatively the **Sugiyama**(Eiglsperger, et al., 2004) family of layered graph layout algorithms provide the additional benefit of providing a flow-diagram structure with links all focused in the same direction (e.g. top down in Figure 7). There also exist implementations capable of laying out nested (hyper)graphs (see for example Microsoft Automatic Graph Layout https://github.com/Microsoft/automatic-graph-layout).

**The premise of this paper is that the key to reducing the risk of using spreadsheets is increasing their level of abstraction. We contend that this is precisely analogous to redrawing the boundaries of the hypergraph edges representing the cells of the spreadsheet.**

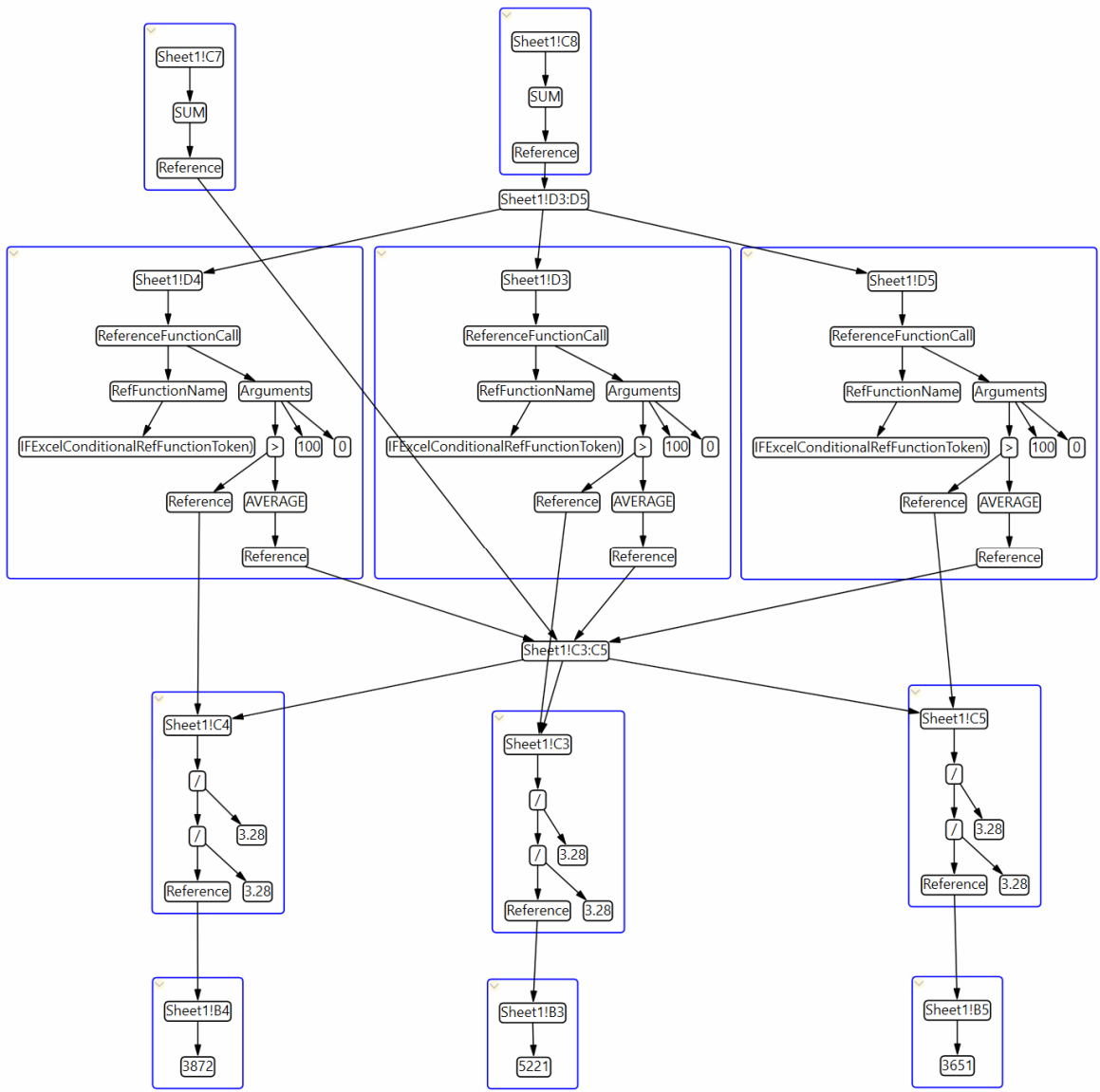

*Figure 7 Hypergraph view of the spreadsheet shown in Figure 1. Note the repeated structure at each level of the graph*

## 8   STRUCTURE DETECTION

By analysing this hypergraph view of the spreadsheet we seek mechanisms for abstracting the spreadsheet by redrawing the boundaries of cells. To identify opportunities for this process we seek to find repeated structure within the graph which may be replaced by a higher level more abstract "cell". In order to do this we explore three abstracting operations:

1) We seek table structures of input values.

2) We seek vector (array) operations which apply the same operations to related data.

3) We seek common sub-expressions within cells which may be factored out to simplify the graph.

### 8.1 Input vector detection

*"{and positioned to reflect its structure}"*

Our objective here is to identify the inputs to a spreadsheet graph which have been entered by the user. Such inputs fall into two general groups, those that are set and used alone such as model **assumptions** like exchange rates and those that are used in lists or tables. We term such larger sets of inputs **measures** which are often related by an index or **dimension** such as time, sales office or product name. Such dimensions may be shared across multiple measures for example sales in different offices over time. Such cross products are common and lead to multidimensional models which can be difficult to model in spreadsheets.

These larger structures are closer to the users mental model and are generally what the users seek, in objective, to manipulate (as appose to a group of loosely connected cells). We seek to identify them so as to abstract them as single logic entities.

To undertake this we consider the subgraphs defined by each cell hyper-edge (these sub-graphs form a tree structure with the cell node at the root) and identify those which a) have no external references and b) who's root node is referenced by at least one (standard) edge. This definition permits us to include simple calculations (=24*60) as inputs in contrast to other approaches. One could further extend this definition to permit references to assumptions as defined above.

Once such candidate input cells are detected we seek to learn from the intuition the user has positioned them with. We do this by using the common Greedy Rectangles algorithms as in (Sarkar, et al., 2018) where a cell is chosen from the list and a vector is grown rightward if the cells exist in the list, the vector is then grown downward if all the cells in that new row exist in the list. Once a vector may grow no more the cells are removed from the list and the process is repeated with a new seed.

We suggest one improvement in this approach which is to permit the vector to grow over blank cells (which are not in the list) provided conditions are met which avoid the elision of adjacent but spatially disjoint vectors. For example should the first row of a table contain a blank cell then the table may truncated. Alternatively one might consider merged cells above or right which likely represent table legend to permit to enable vectors to bridge blank spaces and grow beyond them. Continuous columns or rows of textual legend may also serve the same purpose. Such an extension enables sparse tables to be identified. Further we note that the choice of initial search direction (leftward or downward) does place limitations on the size of vectors which are identified in some cases. Computing in both directions and removing the largest discovered vector may be preferable.

### 8.2 Isomorphic Subtree Vector detection

Using the premise that cells that manipulate data approximately the same way are likely related we seek to identify isomorphically identical cell subgraphs. Two sets are said to be isomorphic if there is a "one to one" mapping function between them which preserves relations between their elements. In our hypergraph we are seeking to identify group of hypergraph edges which demark sufficiently identical sub-graphs that we may consider replacing them with a single "master" version which operates on more data. This abstracting may be considered redrawing the boundary of a cell to include all similar copies. The new "cell" with larger cells walls would, we hope be easier to interact with and manipulate than a large group of cells.

To detect vector operations we first start with the subgraphs formed from the nodes of each of the cell's hyper-edge. These form a strict tree structure. We identify these graphs by walking the graph from the root cell node until other cell nodes are identified (we include cell nodes but walk no further). This permits us to walk over and include Range and Named ranges such that a reference to A1:A2 and A1:A3 may be distinguished by the number of children of the Range node.

The nodes across all of these trees are split into equivalence classes based on their type which is assigned a prime number as an identifier. The rules for defining whether two nodes are within an equivalence classes permit interesting variation in the applicability and success of what this approach will find. For the objective of identifying structurally isomorphic "cells" in terms of the formula they represent we assign all function and operator nodes to their own classes (+, -, sum, if,…), all textual inputs are grouped into one class, numeric inputs also form their own single class. Finally all reference nodes (to a single cell) are placed in the same class.

Having defined such classes each "cell" tree is then walked in Post order (visiting children prior to the node) so that an equivalence class value may be set for each node. This is done by multiplying together the equivalence classes of each of a node's children with its own equivalence class identifier. Since all identifiers are prime numbers we are guaranteed that, up to the permutation of its children, the identifier generated for a node will be unique. Of course this permutation may affect the outcome of a formula (e.g. 0/1 vs 1/0) and should be checked as part of result validation.

Once classes have been calculated for all nodes in each sub-tree, vector detection may proceed by grouping cells by the equivalence class of their root node. This signifies they root an isomorphic subtree. Since we wish to learn from the intuition the user imparted to the spreadsheet by positioning the cells to reflect the structure of their world we seek to create spatially adjacent vectors within each isomorphic group of cells. This is achieved by running the modified greedy rectangles algorithm discussed in the previous section. We further suggest that one might consider accepting vectors with constant gap or offset (e.g. every other column) as this is a common pattern.

Unfortunately it is not sufficient to identify simply **Structurally Isomorphic** cells, further tests must be made:

i. **Constant Isomorphism** – isomorphic formulas must contain identical constants (textual and numeric). This is achieved by walking each tree in the same order (e.g. post order) and extracting a list of constants. The list of cells may then be partitioned by testing pairwise equality on their lists of constants.

ii. **Reference Isomorphism** – Isomorphic cell graphs should make reference in a isomorphic manner, that is there should be an equivalent reference in B for each of the references in A. The obviously isomorphic case to identify is if each graph references the same cell (or same set of cells), such as a shared assumption. However if all cell graphs reference a cell to their left we would still seek to identify an isomorphism by identifying this mapping. This is done by extracting the relative offset of the cell reference. Thus in a similar way to i, one may walk each cell graph in post-order forming a reference list and then partition a group of cell graphs into reference isomorphic subgraphs by considering a pairwise comparison of their reference lists which successes if either A) the cell referenced has the same identifier or B) the same relative offset from the origin cell at the root of the tree. We note that since we walked over the Range and Named range nodes so as to include them in the "cell" subtree they are dealt with by this algorithm without special case.

iii. One might also consider isomorphism by ensuring that cells are referenced in an identical way within the graph, and this may be done via the same method but considering the incoming edges to the root cell node. However experience shows us that this is not an effective means of recovering user intent as vectors are often broken by users "special" cases – such as adding a reference to the 3$^{rd}$ column of a vector to highlight "Sales in March".

Having now grouped cell subgraphs into vectors that meet all of the following isomorphic criteria:

1) **Structural isomorphism** – the same operators in the same order
2) **Colocation** – located in a contiguous cell space
3) **Constant isomorphism** – using the same textual and numeric constants in its calculation
4) **Reference isomorphism** –using isomorphic references within its calculation.

We may redraw the boundaries of these cells to coalesce them into a single larger more abstract "cell" which operates upon vector(s) of data. Such a single data manipulation should prove easier to work with than individually modifying the original group of cells.

One can consider this abstraction process as analogous to extracting and using a function method in a programming language. As in that case our goal is to provide the users with a more abstract and reusable representation of a piece of logic which makes it easier to "link cells by user intent".

|   | A | B | C | D |
|---|---|---|---|---|
| 1 |   |   |   |   |
| 2 | Agent | Sales (sqft) | Sales(m2) | Bonus |
| 3 | Fred | 5221 | 485.29521 | 100 |
| 4 | Dave | 3872 | 359.90482 | 0 |
| 5 | Bob | 3651 | 339.36273 | 0 |
| 6 |   |   |   |   |
| 7 |   | Total Sales | 1184.5628 |   |
| 8 |   | Total Bonus | 100 |   |

*Figure 8 Detected input vector B3:B5 and two sets of isomorphic cell graphs C3:C5 and D3:D5*

To demonstrate this, we have conducted experiments on a sample of 2124 Enron spreadsheets. For these, we report an average of 11% input vector cells out of all non-blank cells. Likewise, we identified an average of 16% vector cells, adding to a total compression rate of 27% on average. Figure 9 plots these three components, showing spreadsheet size measured in non-blank cells on the y axes, as well as the input, vector, and total percentages on the x axes. Naturally, the plots do not include experiments with non-zero vector components.

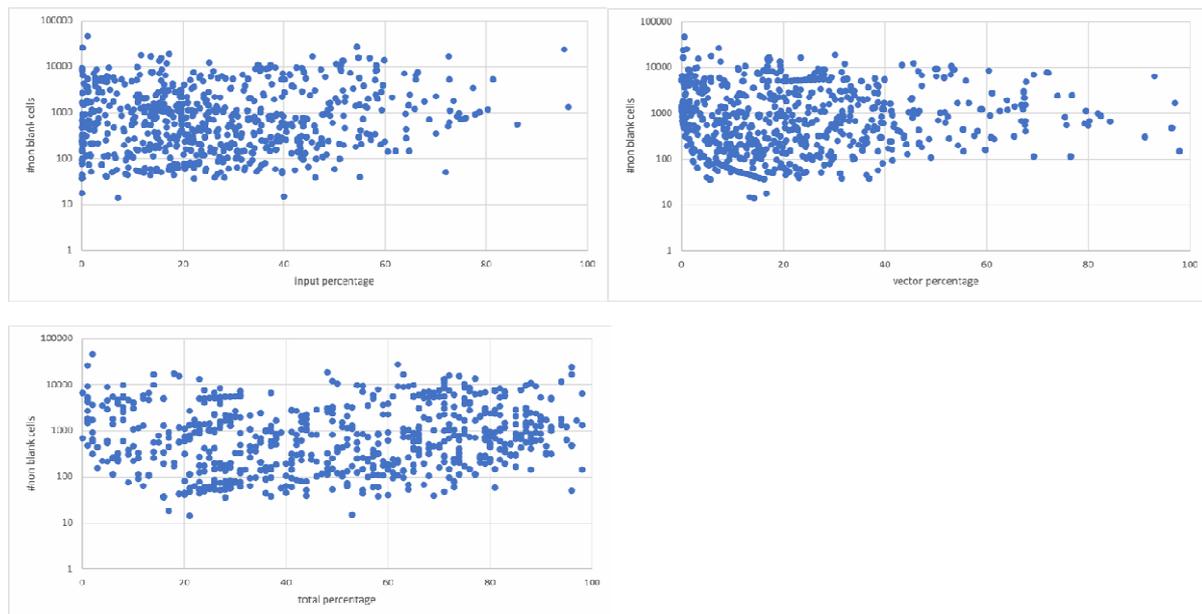

*Figure 9 Plotting the number of vector, input, and total percentages for a sample of the ENRON spreadsheet corpus*

### 8.3 Common Sub-expression Detection

Another redrawing operation for our hypergraphs is to consider the centralisation of duplicated parts of the graph. For example perhaps a calculation is commonly performed – such as conversion from square feet to square meters in the spreadsheet in Figure 1. Alternatively perhaps a common series of formulas are frequently combined to perform a common task. We know this frequently the case from (Middleton & Murphy-Hill, 2016) who studied common function pairings. Extracting these common sub-expressions would further abstract the spreadsheet and reduce the burden placed upon the end user.

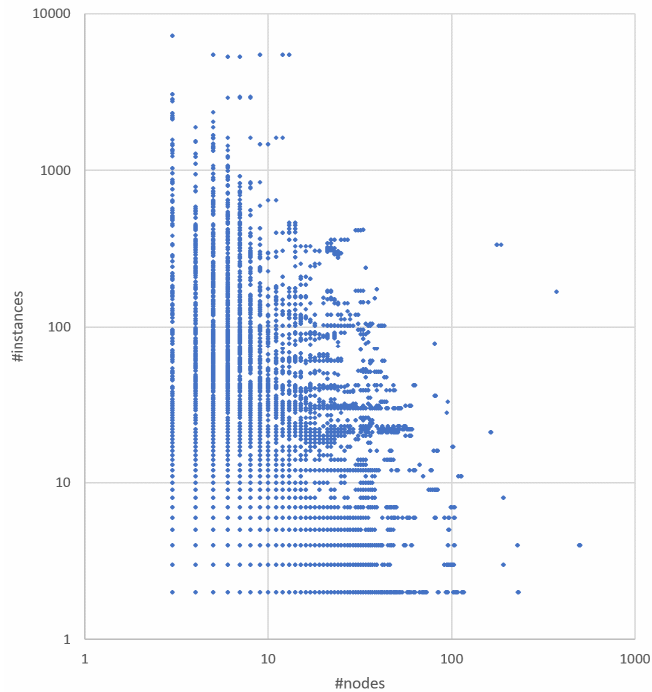

*Figure 10 Plot of the number of structurally isomorphic sub-expressions within a sample of the Enron spreadsheet corpus*

To explore the potential for this approach on a finer grained graph we analysed 694 spreadsheets taken the Enron corpus and sought to identify how many structurally isomorphic subtrees they contained. On average each spreadsheet had 43.4 different subtrees of at least 3 nodes. The average number of nodes within a subtree was 15 and the average number of instances of the subtree per spreadsheet was 90. Figure 10 shows the distribution of the size of each subtree and the number of instances for subtrees of at least 3 nodes.

While these instances would need to be tested for other forms of isomorphism we believe this demonstrates the potential for a common sub-expression analysis to reduce the scale of spreadsheets by factoring out commonality and permitting end user programmers to follow the DRY (don't repeat yourself) principle common in programming.

## 9   A WORLD WITH BIGGER CELLS

Our grant partner Filigree Technologies is building a commercial data modelling platform 'Wire' based on the idea of working at a higher abstraction level than available in spreadsheets. The model logic is presented, and interacted with, in a graph based dataflow environment, which processes the multi-dimensional measure structures discussed earlier. The data flow environment may be seen in Figure 11.

By allowing the user to operate at a higher abstraction level, where the formulae expressions ('modl') embody the conceptual meaning of the processing applied, models are significantly reduced in perceived complexity, resulting in fewer errors and easier maintenance plus collaboration. In internal tests models can be expressed in around 2-5% of the formulae utilised for equivalent spreadsheet models, with a corresponding reducing in boilerplate errors.

Separating data structure, view structure and logical data flow additionally offers considerable benefits in:
- reusing logic as templates which can be simply reapplied to new situations, without the need for repetitive manual restructuring seen in spreadsheets
- handling live data sources, where incoming data structure potentially varies over time, to allow model results to stay up to the minute
- exploring the resultant model to ascertain response to volatility in input assumptions, so as to make the most informed decisions

Utilising the research presented has allowed Excel spreadsheets to be imported into the Wire environment via an intermediate ModL representation, enabling spreadsheet models to be translated to a higher abstraction level, for subsequent maintenance, modification and exploration. As an example in Figures 11 and 12, we introduce a ModL implementation of the spreadsheet in Figure 1. The number of cells in the Excel model is 20 for 3 sales staff (8+N*4) the corresponding ModL graph contains 12 nodes. Each of which has a bigger boundary than a spreadsheet cell. Moreover consider the relationship between the number of cells and the numbers of sales staff in the spreadsheet it is 8+N*4. In ModL the number of graph nodes remains constant at 12 and requires no manual intervention to scale – whereas the sales and bonus totals would both need manual, error prone, adjustment.

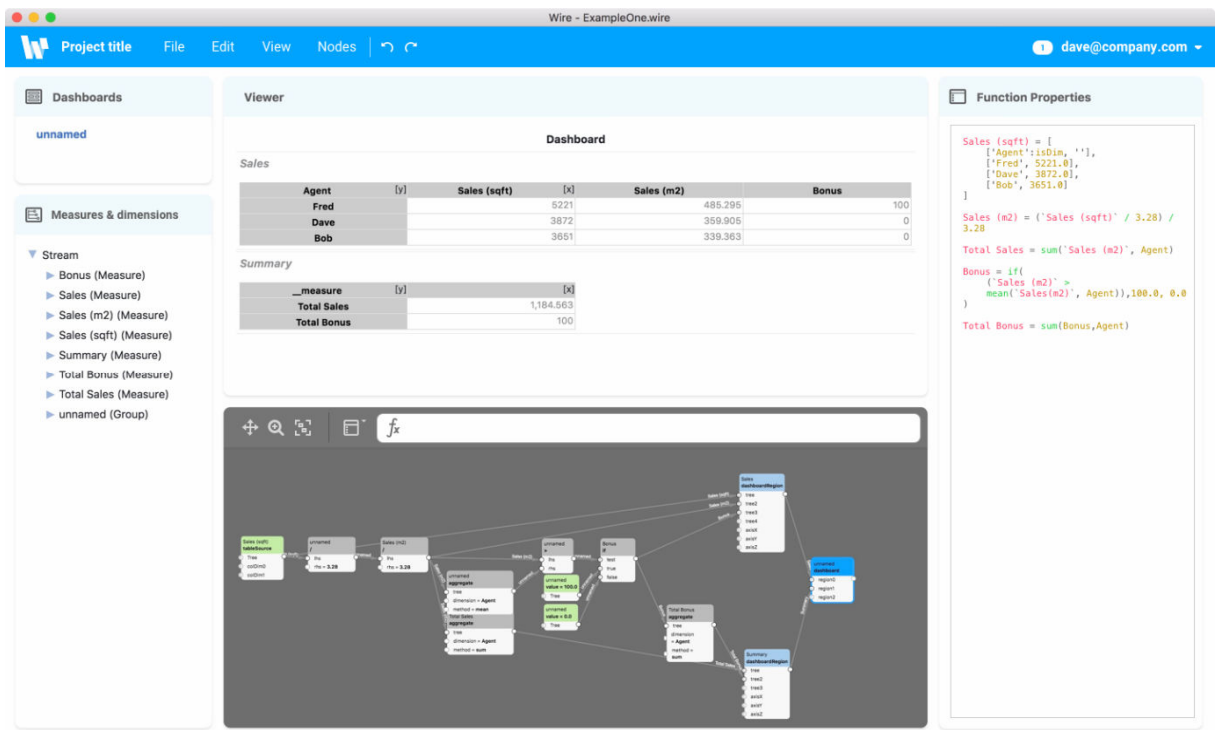

*Figure 11 Translation of the spreadsheet from Figure 1 in Filigree technologis ModL language (right) and shown as a dataflow graph (bottom centre).*

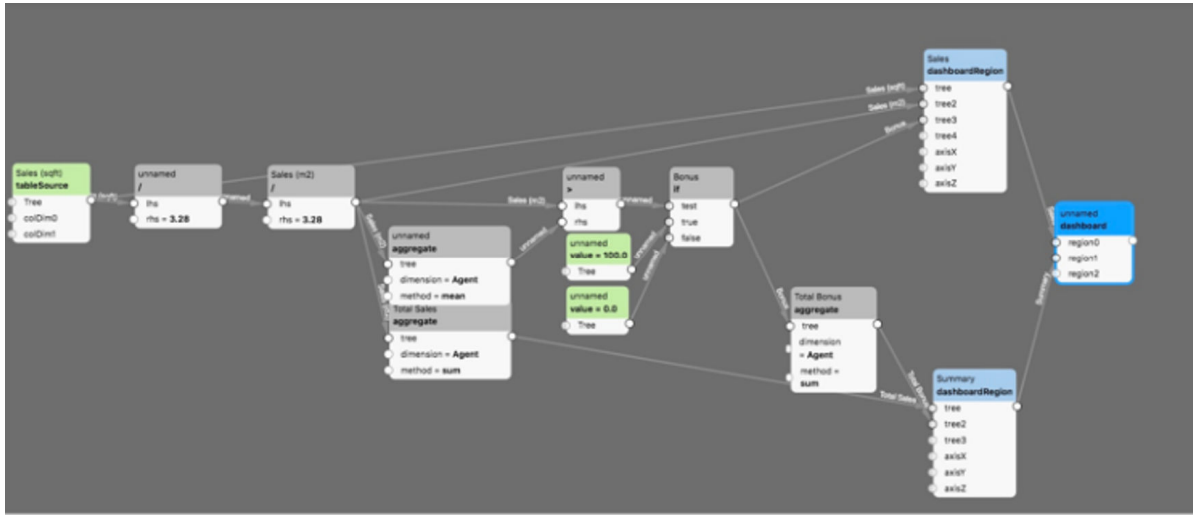

*Figure 12 Calculation graph from Figure 11, implementing the model in Figure 1*

## 10 RELATED WORK:

There is a rich history of academic research into spreadsheets and this work draws heavily from it. Most notably this builds our Extraction and Analysis methodology for spreadsheets introduced in (Birch, et al., 2013). There is a wide literature exploring the prevalence and nature of errors in spreadsheets. A good survey is provided in (Panko, 1998). The spreadsheet grammar used as the basis of our work is introduced by (Aivaloglou, et al., 2015). Other areas of related literature are as follows:

### 10.1 Spreadsheet Corpuses

A number of spreadsheet corpuses are available to researchers. FUSE (Barik, et al., 2015), EUSES (Fisher & Rothermel, 2005) and ENRON (Hermans & Murphy-Hill, 2015) being the most notable. We would commend the comparison in (Jansen, 2015) which inspired our choice of the ENRON corpus as test matter for this paper.

### 10.2     Extraction of Structure

A number of works have sought to extract structure from spreadsheets through 2d search algorithms exploiting the "structured to model their world" motif of cells

- ➢ (Pertti & Sajaniemi, 1991) is an example of early work on extraction of the tree structure formed by spreadsheet cells and their references a simple categorisation of such references is then proposed.

- ➢ (Mittermeir & Clermont, 2002) seek to identify high level structure and introduce a concept of copy equivalence and structural equivalence which is equivalent to our isomorphic AST graph. The author would commend study of (Clermont, 2003) for many detailed insights into structure extraction.

- ➢ (Hermans, et al., 2010) go one step further and seek to extract class diagrams which resemble those familiar to software engineers

- ➢ The visualization of these structures is generally represented within the spreadsheet such as in (Hodnigg & Pinzger, 2015)

- ➢ On a more fine-grained basis (Middleton & Murphy-Hill, 2016) seek to understand frequently combined functions in a visual way.

- ➢ (Eberius, et al., 2013) takes a higher level view and seeks to extract relational data from spreadsheet tables through a process of normalisation more commonly found in the database community.
- ➢ More recently machine learning has been used for the detection of spreadsheet tables(Koci, et al., 2016) this method has been shown to be effective(Koci, et al., 2017) since it is able to deal with the noise generated by inconsistent labels and formulas introduced by spreadsheet users.

In addition there are several spreadsheet audit tools which include the ability to colour tools that visualise the blocks of identical formulas, these include FastExcel (http://www.decisionmodels.com/fastexcel.htm) and XLTest (http://www.sysmod.com/xltest/index.htm).

## 10.3 Spreadsheet Refactoring

In recent years a few works have applied the software engineering process of code refactoring to spreadsheets. This is where code is transformed so that the same functionality is expressed in a "better" way, perhaps using a different more concise code concept – such as a for loop or ternary expression. This refactoring process is in essence a transformation of a spreadsheets underlying AST graph. Indeed the redrawing of cell boundaries is precisely a form of refactoring and a quite general one. We believe that several recent works could be implemented through the process we describe.

- ➢ (Badame & Dig, 2012) introduce 7 refactoring tools for end users such as the introduction of cell names and the extraction of constants as well as checking whether a range of cells have constant formulas which is precisely analogous to our isomorphism check.
- ➢ (Zhang, et al., 2018) use an AST driven approach to specifically consider the simplification of the very common nested IF statement.

Most of these refactoring techniques work within the existing cell walls in that they always refactor to a cell compatible format, we believe that further value could be given by permitting refactoring "outside" the cell walls.

## 10.4 Visualisation

The transformation of spreadsheets into a different visual formats for easier comprehension has a rich and inventive history:

- ➢ (Igarashi, et al., 1998) introduce the now common Dataflow graph and permit it to be edited directly by the end users. Unusually the animation of these graphs and spreadsheet annotation (colouring and arrows) is discussed, primarily through the step by step reveal of the "march of computation" to provide a "narrative expression". It would be interesting to evaluate the efficacy of animation in aiding comprehension as to the authors knowledge this has not been considered before. The force directed algorithms discussed in our paper provide an interesting opportunity for animation to transition between the grid structure and a data flow driven graph layout.
- ➢ (Hermans, et al., 2011) uses a concept of a levelled data flow diagram to represent different conceptual levels of a spreadsheet reflecting the workbooks of the model. This is extended in (Hermans, et al., 2011) to provide a visualisation tool for aiding spreadsheet comprehension.
- ➢ (Shiozawa, et al., 1999) provides an introduction to the use of 3d to visualise different layers of spreadsheets which are taken as the data, formula and reference abstractions.

- ➢ (Kankuzi & Ayalew, 2008) visualise a cell level reference graph using tree view and compound fisheye graph layouts generated using a Markov Clustering Layout algorithm to group cells. It would be interesting to compare this algorithm to its force-directed cousins outlined in the paper.
- ➢ The efficacy of spreadsheet visualisation is an important question and the authors would commend the reader to consider (Goswami, et al., 2008)

**10.5 New Spreadsheet tools**

Over the years many tools have been prese10nted with a view to replacing the spreadsheet. Some such as (Litt, 2018) seek to augment implementations such as Excel with new functionality such as the ability to manage multi-dimensional data. It is speculation to consider the future of the spreadsheet. It is surprising the technology is now approaching its 40$^{th}$ anniversary a collection of the challenges facing spreadsheets may be found in (Birch, et al., 2017)

**10 CONCLUSIONS AND FURTHER WORK:**

In this paper we have argued that one of the central challenges of the spreadsheet and the cause of much of its error-proneness is its low level of abstraction centred around the concept of a cell. We believe there is nothing inherently wrong with the concept of a cell which define as

*"a data container or data manipulator*

*linked by user-intent to model their world*

*and positioned to reflect its structure"*

However we believe current implementations are constrained by their low level nature in that they only hold simple scalar values or the means to generate them. The lack within modern spreadsheets of higher level abstractions which can be manipulated as easily as cells leads to large volumes of cells which in aggregate represent a high level concept. Such groups of cells are generally only associated by their spatial positioning to reflect a higher level structure through use of a grid.

What constrains spreadsheets to their grid? We believe the grid is now primarily used to enable end users to reflect the structure of their world and its data by positioning cells adjacently. If the concept of a cell is expanded to permit it to contain higher level structures such as lists and data tables then the need for this constraining structure is reduced. Instead the natural dataflow with which users link the cell to model their world can be brought to the fore through graph-based dataflow. We believe this is the future for spreadsheets.

We have shown that the transformation of a spreadsheet into a fine-grained operator graph permits us to consider cells as hypergraph edges grouping a set of operators. We have discussed how visualisation of this graph structure may be achieved to highlight the hidden calculation structure of the spreadsheet and introduced a graph layout algorithm which permits the interactive blending of the layout between the calculation structure and the user imposed grid structure.

Within this hypergraph many spreadsheet refactoring's can then be seen as the redrawing of the boundaries of this group. For example the introduction of a named cell or the replacement of a common subexpression.

Further we have explored spreadsheet vectorisation through this motif and shown that through the use of subtree isomorphisms we can identify numerous opportunities in real-world spreadsheets for vectorisation to reduce the size and complexity of spreadsheets and their volume of formulas.

The utility of such an approach however depends upon the adoption of a modelling framework with "larger cell walls", that is where cells may contain more than scalar values and where data flow is a primary user facing concern.

We believe that there remains much further work to done to help end user programmers and the spreadsheets they create enter the age of big data successfully. We see three streams of future work.

Firstly there is much scope for the introduction of new spreadsheet refactoring's using hypergraph redrawing and subtree isomorphisms and the reconstruing of existing refactoring's into this formalism.

Secondly there is a need to develop new spreadsheet inspired formalisms with larger cell walls which support a higher level of abstraction without alienating accomplished end user spreadsheet programmers.

Finally we have shown the potential for the vectorisation of spreadsheets through the identification of isomorphic operations, this paves the way for rapid performance increase and for spreadsheets to process dramatically larger volumes of data in the age of big data.

**Acknowledgements:**

The work has been supported by InnovateUK under grant reference 104141 and through EPSRC Platform Grant EP/P010040/1. This work would not have been possible without the contribution of Filigree technologies who are actively "enlarging the cell walls".